\documentclass[pdfusetitle,aps,prl,twocolumn,nofootinbib,preprintnumbers,superscriptaddress,groupedaddress]{revtex4-1}
\pdfoutput=1
 \usepackage{hyperref}
\hypersetup{colorlinks=true,linkcolor=magenta,anchorcolor=green,citecolor=cyan,filecolor=black,menucolor=black,urlcolor=black}
\usepackage{graphicx} 
\usepackage[table]{xcolor}
\usepackage{graphicx}
\usepackage{epstopdf}
\usepackage[utf8]{inputenc} 
\usepackage{amsmath,amssymb}
\usepackage{soul}
\usepackage{cancel}

\newcommand{\be}{\begin{equation}}
\newcommand{\ee}{\end{equation}}
\newcommand{\bea}{\begin{eqnarray}}
\newcommand{\eea}{\end{eqnarray}}

\newcommand{\bdm}{\begin{displaymath}}
\newcommand{\edm}{\end{displaymath}}

\definecolor{bubbles}{rgb}{0.91, 1.0, 1.0}
\definecolor{aquamarine}{rgb}{0.5, 1.0, 0.83}
\definecolor{bubblegum}{rgb}{0.99, 0.76, 0.8}
\definecolor{bluebell}{rgb}{0.64, 0.64, 0.82}
\definecolor{dollarbill}{rgb}{0.72, 0.93, 0.6}

\usepackage{tikz,fp}
\usetikzlibrary{external,fixedpointarithmetic,decorations.pathmorphing,positioning,calc,fadings,decorations.markings,decorations.pathreplacing,patterns,arrows}

\tikzset{
  label distance=-3pt,
  >=stealth,
  inner sep=1.5pt,
  boson/.style={
    decoration={snake, segment length=2mm, amplitude=0.5mm},
    decorate,
  },
  scalar/.style={
    dashed
  }
}

\begin{document}
\preprint{IPHT-t24/010}
\title{Schwarzschild metric from Scattering Amplitudes to all orders
  in $G_N$}
\author{Stavros~Mougiakakos}
\affiliation{Laboratoire Univers et Th\'eories, Observatoire de Paris, Universit\'e PSL, Universit\'e Paris Cit\'e, CNRS, F-92190 Meudon, France}
\author{Pierre~Vanhove}
\affiliation{Institut de Physique Th\'eorique, Universit\'e  Paris-Saclay, CEA, CNRS, F-91191 Gif-sur-Yvette Cedex, France}

\date{\today}

\begin{abstract}
We apply a formulation of Einstein's general relativity with only
cubic interactions for deriving the metric of a Schwarzschild black
hole to all orders in perturbation theory. This cubic interactions
formulation coupled to effective worldline action of a massive point
particle allows to derive a recursion relation for the form factors of
the off-shell graviton emission current. The unique solution to the
recursion relation leads to the Schwarzschild black-hole solution in
four dimensions. This provides the first derivation of the black hole
metric from a matter source to all orders in perturbation theory from
an amplitude approach.
\end{abstract}

\maketitle

\section{Introduction}
The necessity for high precision analytic computations of gravitational waves emitted from the gravitational two-body scattering in Einstein gravity has led to the development of important theoretical frameworks~\cite{Bjerrum-Bohr:2022blt,Kosower:2022yvp,Buonanno:2022pgc}.  The
remarkable connection between the perturbative quantum
scattering method and the classical post-Minkowskian expansion, as demonstrated in~\cite{Bjerrum-Bohr:2018xdl,Kosower:2018adc}, has
provided analytic expressions up to the fourth post-Minkowskian order
in the spinless
case~\cite{Bjerrum-Bohr:2013bxa,Cheung:2018wkq,Bern:2019nnu,Mogull:2020sak,Kalin:2020mvi,Kalin:2020fhe,Dlapa:2021npj,Bjerrum-Bohr:2021din,Bern:2021dqo,Bern:2021yeh,Herrmann:2021tct,DiVecchia:2021bdo,Dlapa:2022lmu,Kalin:2022hph,Damgaard:2023ttc,Georgoudis:2023eke,Georgoudis:2024pdz,Bini:2024rsy},
with the conservative sector result at the fifth
post-Minkowskian order~\cite{Driesse:2024xad}. These results are in impressive agreement with those obtained from numerical general relativity~\cite{Khalil:2022ylj,Rettegno:2023ghr}. A variety of amplitude methods have been employed to derive the  space-time metric of a black hole induced by a massive object in perturbation from scattering amplitudes techniques~\cite{Boulware:1968zz,Duff:1973zz,Bjerrum-Bohr:2002fji,Neill:2013wsa,Jakobsen:2020ksu,Mougiakakos:2020laz,DOnofrio:2022cvn}. 
The intricacy of the computational process precluded the derivation of the complete metric to all orders in perturbation. Recurrence relations between the off-shell graviton emission from a massive source was obtained in~\cite{Damgaard:2024fqj} using the perturbiner approach---which encode the classical equations of motion---leading to the complete Schwarzschild metric.

In this work, we show how to derive the Schwarzschild metric to all orders in the gravitational
coupling from a scattering amplitude computation. The approach uses
the reformulation of the Einstein-Hilbert action with only cubic interactions~\cite{Cheung:2017kzx} coupled to a massive source within a worldline EFT formalism.
The graviton or auxiliary field off-shell emission from the
massive source are only given by multiloop triangle Feynman diagrams which is
expected from general arguments about the emergence of the classical
contribution from the quantum amplitude~\cite{Bjerrum-Bohr:2018xdl}. The worldline construction directly provides the velocity on the massive lines~\cite{Bjerrum-Bohr:2021din,Damgaard:2023vnx} and reduce the computation to the multiloop massless bubble master integrals of~\cite{Mougiakakos:2020laz}.
 The restriction to cubic interactions results to a crucial simplification that at each order in perturbation 
these multiloop triangle diagrams are represented only by binary tree graphs.  This binary tree structure implies a recursive  relation between the various orders in perturbation. 
 The solution to this recursion
relation is unique and is given by the Schwarschild solution in four dimensions.
Remarkably  the derivation of the Schwarzschild metric in this work does not require any
regularisation contrary to the previous
perturbative
computations~\cite{Duff:1973zz,Mougiakakos:2020laz,DOnofrio:2022cvn}.
The freedom from the harmonic gauge
condition~\cite{Fromholz:2013hka} allowed absorbing the short distance
singularities~\cite{Mougiakakos:2020laz} arising in the amplitude computation removing all
possible ambiguities from the quantum computation, rendering the computation finite and consistent albeit limited to a finite order in perturbation. The new ingredient
in this work is the use of a different set of degrees of
freedom (DOF) given by the perturbative expansion of the \emph{gothic inverse metric}
in~\eqref{e:pertgoth} and the auxiliary field~\eqref{e:Adef}. This set of variables  fix all the remaining
freedom from the harmonig gauge condition, leading to a unique
solution given by the Schwarschild metric to all orders.

\section{Cubic formulation of General Relativity}

In this section, we will briefly present the cubic formulation of GR
as initially introduced in~\cite{Cheung:2017kzx}. We will use the
mostly positive metric signature $(-,+,\dots,+)$. The
Einstein-Hilbert action $\mathcal{L}_{EH}=\sqrt{-g} \mathcal R(g)/(16  \pi G_N )$  can be recast in the form

\begin{equation}
 16  \pi G_N \mathcal{L}_{EH}=-\left(A^a_{bc}A^b_{ad}-\frac{1}{D-1}A^a_{ac}A^b_{bd}\right)\mathfrak{g}^{cd}+A^a_{bc}\partial_a\mathfrak{g}^{bc},
\end{equation}
where   $G_N$ is
Newton's constant, $A^a_{bc}=A^a_{cb}$ is an auxiliary field
and $\mathfrak{g}^{ab}=\sqrt{-g}g^{ab}$ is the \textit{gothic inverse metric}
used to lower and raise the indices
$A^a_{bc}\mathfrak{g}^{bb'}\mathfrak g^{cc'}=A^{ab'c'}$. The corresponding classical equations of motion are

  \begin{eqnarray}
\frac{\delta\mathcal{L}_{EH}}{\delta\mathfrak{g}^{ab}}&=&-A^c_{da}A^d_{cb}+\frac{1}{D-1}A^c_{ca}A^d_{db}-\partial_cA^c_{ab}=0,\\
\frac{\delta\mathcal{L}_{EH}}{\delta A^a_{bc}}&=&\left(A^{(b}_{ad}-\frac{1}{D-1}A^e_{ed}\delta_a^{(b}\right)\mathfrak{g}^{c)d}-\partial_a\mathfrak{g}^{bc}=0,
\end{eqnarray}
 and the second equation can be inverted as
 \begin{equation}\label{e:Adef}
 A^a_{bc}=\frac{1}{2}\left({\mathfrak{g}}_{d(b}\partial_{c)}\mathfrak{g}^{ad}+\mathfrak{g}^{ad}\partial_d\mathfrak{g}^{lm}\left(\frac{ {\mathfrak{g}}_{lm} {\mathfrak{g}}_{bc}}{D-2}-\frac{ {\mathfrak{g}}_{l(b} {\mathfrak{g}}_{c)m}}{2}\right)\right).
 \end{equation}
From now on we will work perturbatively by expanding the \emph{gothic inverse metric}
near flat space 
\begin{equation}\label{e:pertgoth}\mathfrak{g}^{ab}=\sqrt{-g}g^{ab}=\eta^{ab}-
\sqrt{32\pi G_N}  h^{ab}\end{equation}
and the indices are raised and lowered with the flat metric as usual.
Additionally, we shift the auxiliary field as
\begin{equation}\label{shift}
A^a_{bc}\rightarrow A^a_{bc}-\frac{\eta^{ad}}{2}\left(\partial_{(b}h_{c)d}+\frac{\eta_{bc}\partial_d h}{D-2}-\partial_d h_{bc}\right),
\end{equation}
in order to decouple the quadratic interactions between the graviton $h_{ab}$ and the auxiliary field $A^a_{bc}$, since our aim is to work in a perturbative QFT setup.
Finally, we will add the gauge fixing term 
\begin{equation}
\mathcal{L}_{GF}=-{1\over 32\pi G_N}\ \eta_{cd}\partial_a\mathfrak{g}^{ac}\partial_b\mathfrak{g}^{bd},
\end{equation}
which corresponds to the harmonic gauge condition $\partial_a\mathfrak{g}^{ab}=0$ for the \textit{gothic inverse metric}. 
Upon gauge fixing, the Lagrangian becomes
$\mathcal{L}=\mathcal{L}_{EH}+\mathcal{L}_{GF}$ 
containing only cubic self interactions. The corresponding Feynman
rules were derived in~\cite{Cheung:2017kzx}. These Feynman rules adapted to
our conventions are given in the Appendix. The above cubic formulation of GR is crucial to derive an all order result for the off-shell currents as it will be shown in the next section.\\

We provide here for later convenience the exact classical solutions in
$D=4$ in the harmonic gauge~\cite{Fromholz:2013hka} for the \emph{gothic inverse metric} 
perturbation, using $\rho=G_Nm/r$, the spatial unit vector $n^\mu:=(0,x_1/r,\dots,x_{D-1}/r)$
with $r^2=x_1^2+\cdots+x_{D-1}^2$,
\begin{equation}\label{e:hresult.class}
\sqrt{32\pi
  G_N}h^{\mu\nu}=\left(-1+\frac{(1+\rho)^3}{1-\rho}\right)\delta^{\mu}_0\delta^{\nu}_0+\rho^2n^\mu
n^\nu,
\end{equation}
and the auxiliary field perturbation, using~\eqref{e:Adef} and~\eqref{shift}
\begin{multline}\label{e:Aresult.class}
\sqrt{32\pi G_N}A^{a\ (n)}_{bc}(\textbf{x})=\cr\frac{\rho}{r}\Bigg[\frac{(2-\rho)}{(1-\rho^2)}\left(-1+\frac{(1+\rho)^3}{1-\rho}\right)\delta^a_0n_{(b}\delta_{c)}^0\cr
+\left(-1+\frac{2}{(1-\rho)^2}-\frac{1-\rho}{(1+\rho)^3}\right)n^a\delta_b^0\delta_c^0\cr
                              +2\left(-1+\frac{1}{(1-\rho)^2}\right)n^a\delta_{bj}\delta_{ck}\cr
+\rho\left(-2+\frac{1}{1+\rho}+\frac{1}{1-\rho^2}\right)n^an_bn_c
                            \Bigg].
\end{multline}

\section{Off-shell currents from Worldline Formalism}

We derive the off-shell three-point functions, for the graviton and the
auxiliary field, via a worldline EFT to all orders in $G_N$~\cite{Goldberger:2004jt,Porto:2016pyg}. In order to proceed, we express the point-particle wordline action
\begin{equation}
\mathcal{L}_{p.p.}=-\frac{m}{2}\int d\tau\
                      \left(e^{-1} \, g^{\mu\nu}v_{\mu}v_{\nu}+e\right)
\end{equation}
using the new DOF $g^{\mu\nu}=\mathfrak{g}^{\mu\nu}/(\sqrt{-\mathfrak{g}})^{2/(D-2)}$ and fix the einbein to $e=1$ so that 
\begin{equation}\label{e:Lpp}
\mathcal{L}_{p.p.}=-\frac{m}{2}\int d\tau\ \left(\frac{\mathfrak{g}^{\mu\nu}v_{\mu}v_{\nu}}{\left(\sqrt{-\mathfrak{g}}\right)^{\frac{2}{D-2}}}+1\right),
\end{equation}
which has the perturbative expansion to the first order
\begin{multline}\label{e:Lpp}
\mathcal{L}_{p.p.}=-\frac{m}{2}\int d\tau\ \Big[1+v^2-\sqrt{32\pi
    G_N}\, h^{\mu\nu}\,\cr
\times\left(v_{\mu}v_{\nu}-v^2\frac{\eta_{\mu\nu}}{D-2}\right)+\mathcal{O}(G_N)\Big].
\end{multline}
Working in the rest frame of the particle where $v_{\mu}=(-1,0,0,0)$ and $v^2=-1$, the corresponding Feynman rule for the one-graviton emission from the worldline is
\begin{equation}\label{e:tdef}
t^{\alpha\beta}=\sqrt{32\pi
  G_N}\frac{i m}{2}\left(\delta^{\alpha}_0\delta^{\beta}_0+\frac{\eta^{\alpha\beta}}{D-2}\right),
\end{equation}
and is the only worldline vertex that we will need.

In the cubic formulation the projector of the graviton propagator in eq.~\eqref{e:PropG} does not depend on the space-time dimension $D$, and it has the usual form for $D=4$ dimensions~\cite{Veltman:1975vx}. In general dimension the vertices derived from~\eqref{e:Lpp} project on the transverse DOF which ensure that  the  harmonic condition $k^{\mu}h_{\mu\nu}(\textbf{x})=0$  is satisfied.\\

Furthermore, at each order in perturbation in $G_N$  the  graviton and auxiliary field emission 
\begin{eqnarray}
\sqrt{32\pi G_N}h_{\mu\nu}^{(n)}(\textbf{x})=&\!\!\!\displaystyle{\int} d^{D-1}x\ e^{i\textbf{k}\cdot\textbf{x}}\ J^{(n)}_{\mu\nu}(\textbf{k}),\\
\sqrt{32\pi G_N}A^{a\ (n)}_{bc}(\textbf{x})=&\!\!\!\displaystyle{\int} d^{D-1}x\ e^{i\textbf{k}\cdot\textbf{x}}\ Y^{a\ (n)}_{bc}(\textbf{k}),
\end{eqnarray}
are parametrized by  eight scalar dimension dependent \emph{form factors}
\begin{equation}\boldsymbol{\mathbf{\chi}}^{(n)}(D)=(\chi_1^{(n)},\dots,\chi_8^{(n)})\end{equation}
entering   the off-shell current for the 1-graviton emission 
\begin{equation}\label{e:Jansatz}
J^{(n)}_{\mu\nu}(\textbf{k})=\rho(|\textbf{k}|,D,n) \,\left(\chi_1^{(n)}\delta_{\mu}^0\delta_{\nu}^0+\chi_2^{(n)}\eta_{\mu\nu}+\chi_3^{(n)}\frac{k_{\mu}k_{\nu}}{\textbf{k}^2}\right),
\end{equation}
and the off-shell current for the auxiliary field emission
\begin{multline}\label{e:Yansatz}
Y^{a\ (n)}_{bc}(\textbf{k})=-i \rho(|\textbf{k}|,D,n)
\,\Big(k_{(b}\left(\chi_7^{(n)}\delta_{c)}^0\delta_0^a+\chi_8^{(n)}\delta^{a}_{c)}\right)
\cr+
k^a\left(\chi_4^{(n)}\delta_b^0\delta_c^0+\chi_5^{(n)}\eta_{bc}+\chi_6^{(n)}\frac{k_bk_c}{\textbf{k}^2}\right)
\Big).
\end{multline}
The order parameter in momentum space
\begin{equation}
  \label{e:rhodef}
  \rho(|\textbf{k}|,D,n)=\frac{\Gamma\left(\frac{2-(D-3)(n-1)}{2}\right)}{\Gamma\left(\frac{n(D-3)}{2}\right)}\,\frac{\left(\Gamma\left(\frac{D-3}{2}\right)
    G_Nm\right)^n}{(|\textbf{k}|/(2\sqrt{\pi}))^{2-(D-3)(n-1)}},
\end{equation}
which Fourier transforms to
\begin{equation}
  \int {d^{D-1}k\over (2\pi)^{D-1}}e^{i\textbf{k}\cdot \textbf{r}} \rho(|\textbf{k}|,D,n)\, = \rho(r,D)^n,
\end{equation}
with the coordinate space order parameter~\cite{Mougiakakos:2020laz}
\begin{equation}
  \rho(r,D)= {\Gamma\left(D-3\over 2\right)\over \pi^{D-3\over2}}
  {G_Nm\over r^{D-3}}
\end{equation}
which evaluates to the Schwarzschild parameter in four dimensions $\rho(r,4)=\rho$.
Notice that the transversality of the graviton ${k}^\mu
h_{\mu\nu}=0$ implies that
$\chi_2^{(n)}+\chi_3^{(n)}=0$ for all $n$. Starting with these two independent
form factors we will derive this relation as a consequence of the
gauge invariance of the computation.

\subsection{Tree-level order}

Because only the graviton couples directly to the matter line, at
tree-level there is only the 1-graviton emission contribution (we
refer to the Appendix for notations and conventions on the Feynman rules)
\begin{equation}
J_{\mu\nu}^{(1)}=
    \begin{tikzpicture}[baseline={([yshift=-.5ex]current bounding box.center)},scale=1]
      \draw[boson] (0,-0.5) -- (0.0,-1.5);
     \draw [fill] (0,-1.5) circle [radius=2pt];
   \end{tikzpicture}
    =\sqrt{32\pi
      G_N}\mathcal{P}^h_{\alpha\beta\mu\nu}t^{\alpha\beta}=\rho(|\textbf{k}|,D,1)
    4\delta_{\mu}^0\delta_{\nu}^0,
  \end{equation}
and the form factors can be identified as 
\begin{equation}\label{e:val1}
\boldsymbol{\chi}^{(1)}(D)=(4,0,0,0,0,0,0,0).
\end{equation}

\subsection{One-loop order}

At one-loop order we have two contributions, the graviton emission and
the auxiliary field emission. The off-shell graviton emission is given by
\begin{equation}
     J^{(2)}_{\mu\nu}=\frac{1}{2}
    \begin{tikzpicture}[baseline={([yshift=-.5ex]current bounding box.center)},scale=1]
      \draw[boson] (0,-0.5) -- (0.0,0.05);
      \draw[boson] (0,-0.5) -- (-0.5,-1.5);
      \draw[boson] (0,-0.5) -- (0.5,-1.5);
      \draw [fill] (-.5,-1.5) circle [radius=2pt];
         \draw [fill] (0.5,-1.5) circle [radius=2pt];
    \end{tikzpicture},
  \end{equation}
  which reads using the Feynman rules of the Appendix
        \begin{equation}
    J^{(2)}_{\mu\nu}=\frac{\sqrt{32\pi
                        G_N}}{2}\!\!\int \!\!{d^{D-1}q\over (2\pi)^{D-1}}
                      \mathcal{P}^h_{\alpha\beta\kappa\lambda}t^{\alpha\beta}\mathcal{P}^h_{\gamma\delta\rho\sigma}t^{\gamma\delta}V_{h^3}^{\epsilon\zeta\kappa\lambda\rho\sigma}\mathcal{P}^h_{\epsilon\zeta\mu\nu},
\end{equation}
is given by the one-loop bubble contribution which is the master integral
arising at this order~\cite{Mougiakakos:2020laz} 
\begin{multline}
  J^{(2)}_{\mu\nu}=
                      \int   {d^{D-1}q\over (2\pi)^{D-1}}  \frac{8\pi^2 \left(G_Nm\ \chi_1^{(1)}\right)^2}{(\textbf{q})^2(\textbf{q}-\textbf{k})^2}\cr
\times                      \left(\left(1-\frac{(D-3)}{4(D-2)^2}\right)\delta_{\mu}^0\delta_{\nu}^0+\frac{(D-3)^2}{4(D-2)^2}\left(\eta_{\mu\nu}-\frac{k_{\mu}k_{\nu}}{|\textbf{k}|^2}\right)\right)
\end{multline}
which evaluates to
\begin{multline}
  J^{(2)}_{\mu\nu}
  =\rho(|\textbf{k}|,D,2)\,
      \frac{\left(\chi_1^{(1)}\right)^2}{2}\Bigg(\left(1-\frac{(D-3)}{4(D-2)^2}\right)\delta_{\mu}^0\delta_{\nu}^0\cr
      +\frac{(D-3)^2}{4(D-2)^2}\left(\eta_{\mu\nu}-\frac{k_{\mu}k_{\nu}}{|\textbf{k}|^2}\right)\Bigg).
    \end{multline}
The off-shell auxiliary field emission is given by
    \begin{eqnarray}
&&Y^{a\ (2)}_{bc}=\frac{1}{2}
    \begin{tikzpicture}[baseline={([yshift=-.5ex]current bounding box.center)},scale=1]
      \draw[scalar] (0,-0.5) -- (0.0,0.05);
      \draw[boson] (0,-0.5) -- (-0.5,-1.5);
      \draw[boson] (0,-0.5) -- (0.5,-1.5);
      \draw [fill] (-.5,-1.5) circle [radius=2pt];
         \draw [fill] (0.5,-1.5) circle [radius=2pt];
    \end{tikzpicture}\\
    &=&\frac{\sqrt{32\pi G_N}}{2}\!\!\int\!\! {d^{D-1}q\over (2\pi)^{D-1}}
        \mathcal{P}^h_{\alpha\beta\kappa\lambda}t^{\alpha\beta}\mathcal{P}^h_{\gamma\delta\rho\sigma}t^{\gamma\delta}V_{h^2A\
        \ \ \  d}^{\kappa\lambda\rho\sigma\ ef}\mathcal{P}^{A\ a \
        d}_{\ \ \  \ bc\ ef},\nonumber
    \end{eqnarray}
  which explicitly reads
      \begin{multline}
Y^{a\ (2)}_{bc}
    =\left(G_Nm\
        \chi_1^{(1)}\right)^2\int {d^{D-1}q\over (2\pi)^{D-1}}\frac{-32\pi^2i}{(\textbf{q})^2(\textbf{q}-\textbf{k})^2}\cr\times\Big(\frac{k^a}{D-2}\Big((2D-5)\delta_b^0\delta_c^0
        +\eta_{bc}\Big)-k_{(b}\delta_{c)}^0\delta_0^a\Big) 
      \end{multline}
      and evaluates to
  \begin{multline}
Y^{a\ (2)}_{bc}
    =-i 2\rho(|\textbf{k}|,D,2)\,
        \left(\chi_1^{(1)}\right)^2\cr
        \times\Big(\frac{k^a}{D-2}\left((2D-5)\delta_b^0\delta_c^0
        +\eta_{bc}\right)-k_{(b}\delta_{c)}^0\delta_0^a\Big) .
    \end{multline}
   From these results we can identify the eight one-loop form factors
    \begin{multline}\label{e:val2}
  \boldsymbol{\chi}^{(2)}(D)=\Big(8-2\frac{(D-3)}{(D-2)^2},\frac{2(D-3)^2}{(D-2)^2},-\frac{2(D-3)^2}{(D-2)^2}, \cr 8-\frac{4}{D-2},\frac{4}{D-2},0,-4,0\Big).
\end{multline}
To make explicit the iteration we use the tree-level graviton emission
$J^{(1)}_{\mu\nu}$ to express the one-loop emissions  [using the
barred notation introduced in Eq.~\eqref{e:BarPV}]
  \begin{equation}
    J^{(2)}_{\mu\nu}=\frac{1}{2}J^{(1)}_{\gamma\delta}J^{(1)}_{\rho\sigma}\bar{V}_{h^3}^{\alpha\beta\gamma\delta\rho\sigma}\bar{\mathcal{P}}^h_{\alpha\beta\mu\nu}
   =
    \begin{tikzpicture}[baseline={([yshift=-.5ex]current bounding box.center)},scale=1]
      \draw[boson] (0,-0.5) -- (0.0,0.05);
      \draw[boson] (0,-0.5) -- (-0.5,-1.5);
      \draw[boson] (0,-0.5) -- (0.5,-1.5);
      \draw [fill] (-.5,-1.5) circle [radius=2pt];
         \draw [fill] (0.5,-1.5) circle [radius=2pt];
      \node[label=below:\small{$J^{(1)}$}] at (-0.5,-1.5) {};
      \node[label=below:\small{$J^{(1)}$}] at (0.5,-1.5) {};
    \end{tikzpicture}
  \end{equation}
and 
\begin{equation}
    Y^{a\ (2)}_{bc}=-\frac{i}{2}J^{(1)}_{\alpha\beta}J^{(1)}_{\gamma\delta}\bar{V}_{h^2A\ \ \ \  d}^{\alpha\beta\gamma\delta\ ef}\bar{\mathcal{P}}^{A\ a \ d}_{\ \ \  \ bc\ ef}= \begin{tikzpicture}[baseline={([yshift=-.5ex]current bounding box.center)},scale=1]
      \draw[scalar] (0,-0.5) -- (0.0,0.05);
      \draw[boson] (0,-0.5) -- (-0.5,-1.5);
      \draw[boson] (0,-0.5) -- (0.5,-1.5);
      \draw [fill] (-.5,-1.5) circle [radius=2pt];
         \draw [fill] (0.5,-1.5) circle [radius=2pt];
      \node[label=below:\small{$J^{(1)}$}] at (-0.5,-1.5) {};
      \node[label=below:\small{$J^{(1)}$}] at (0.5,-1.5) {};
    \end{tikzpicture}.
  \end{equation}

This iterative structure persists to all order in perturbation and is
one of the key ingredients that lead to a recursion relation connecting
the $l$-loop contribution to lower-loop contributions.

\subsection{Two-loop order}

 At  two-loop order the graviton emission  is given by the sum
    of the two diagrams
\begin{equation}
  J^{(3)}_{\mu\nu}
                      =\frac{1}{2} \left(\begin{tikzpicture}[baseline={([yshift=-.5ex]current bounding box.center)},scale=1]
      \draw[boson] (0,-0.5) -- (0.0,0.05);
      \draw[boson] (0,-0.5) -- (-0.5,-1.5);
      \draw[boson] (0,-0.5) -- (0.5,-1.5);
      \draw[boson] (0.2,-1.0) -- (0.0,-1.5);
      \draw [fill] (-.5,-1.5) circle [radius=2pt];
        \draw [fill] (0.0,-1.5) circle [radius=2pt];
         \draw [fill] (0.5,-1.5) circle [radius=2pt];
      \node[label=below:\small{$J^{(1)}$}] at (-0.5,-1.5) {};
      \node[label=below:\small{$J^{(1)}$}] at (0.5,-1.5) {};
       \node[label=below:\small{$J^{(1)}$}] at (0.0,-1.5) {};
    \end{tikzpicture}
    +\begin{tikzpicture}[baseline={([yshift=-.5ex]current bounding box.center)},scale=1]
      \draw[boson] (0,-0.5) -- (0.0,0.05);
      \draw[boson] (0,-0.5) -- (-0.5,-1.5);
      \draw[boson] (0.2,-1.0) -- (0.5,-1.5);
      \draw[boson] (0.2,-1.0) -- (0.0,-1.5);
      \draw[scalar] (0.2,-1.0) -- (0,-0.5);
      \draw [fill] (-.5,-1.5) circle [radius=2pt];
       \draw [fill] (0.0,-1.5) circle [radius=2pt];
         \draw [fill] (0.5,-1.5) circle [radius=2pt];
      \node[label=below:\small{$J^{(1)}$}] at (-0.5,-1.5) {};
      \node[label=below:\small{$J^{(1)}$}] at (0.5,-1.5) {};
       \node[label=below:\small{$J^{(1)}$}] at (0.0,-1.5) {};
\end{tikzpicture}\right)
\end{equation}
where $J^{(1)}$ is the one-graviton emission from the massive
particle. Because of the cubic structure of the theory, this is the complete set of contributing diagrams. This is of great importance for the generalization of the iterative structure to all orders, as the absence of higher-order self-interactions restricts the topologies of the diagrams to the above binary tree structure. It should be noted that in the worldline formalism, the sources are static, thus all the different symmetrizations of the propagators are equivalent, and the total symmetry factor of the diagram is $1/2$. This feature is applicable to all loops, enabling the symmetry factors to be incorporated into the definitions of the lower-order diagrams.

Using the Feynman rules given in the Appendix, this expression can be 
rewritten with respect to the one-loop order off-shell currents as
\begin{equation}\label{e:J3result}
    J^{(3)}_{\mu\nu}=\begin{tikzpicture}[baseline={([yshift=-.5ex]current bounding box.center)},scale=1]
      \draw[boson] (0,-0.5) -- (0.0,0.05);
      \draw[boson] (0,-0.5) -- (-0.5,-1.5);
      \draw[boson] (0,-0.5) -- (0.5,-1.5);
      \draw [fill] (-.5,-1.5) circle [radius=2pt];
         \draw [fill] (0.5,-1.5) circle [radius=2pt];
      \node[label=below:\small{$J^{(1)}$}] at (-0.5,-1.5) {};
      \node[label=below:\small{$J^{(2)}$}] at (0.5,-1.5) {};
    \end{tikzpicture}-
    \begin{tikzpicture}[baseline={([yshift=-.5ex]current bounding box.center)},scale=1]
      \draw[boson] (0,-0.5) -- (0.0,0.05);
      \draw[boson] (0,-0.5) -- (-0.5,-1.5);
      \draw[scalar] (0,-0.5) -- (0.5,-1.5);
      \draw [fill] (-.5,-1.5) circle [radius=2pt];
         \draw [fill] (0.5,-1.5) circle [radius=2pt];
      \node[label=below:\small{$J^{(1)}$}] at (-0.5,-1.5) {};
      \node[label=below:\small{$Y^{(2)}$}] at (0.5,-1.5) {};
    \end{tikzpicture}.
  \end{equation}
Likewise, for the emission of the auxiliary field we have
  \begin{equation}
    Y^{a\ (3)}_{bc}=\frac{1}{2}\left(
    \begin{tikzpicture}[baseline={([yshift=-.5ex]current bounding box.center)},scale=1]
      \draw[scalar] (0,-0.5) -- (0.0,0.05);
      \draw[boson] (0,-0.5) -- (-0.5,-1.5);
      \draw[boson] (0,-0.5) -- (0.5,-1.5);
      \draw[boson] (0.2,-1.0) -- (0.0,-1.5);
      \draw [fill] (-.5,-1.5) circle [radius=2pt];
       \draw [fill] (0.0,-1.5) circle [radius=2pt];
         \draw [fill] (0.5,-1.5) circle [radius=2pt];
      \node[label=below:\small{$J^{(1)}$}] at (-0.5,-1.5) {};
      \node[label=below:\small{$J^{(1)}$}] at (0.5,-1.5) {};
       \node[label=below:\small{$J^{(1)}$}] at (0.0,-1.5) {};
    \end{tikzpicture}
    + \begin{tikzpicture}[baseline={([yshift=-.5ex]current bounding box.center)},scale=1]
      \draw[scalar] (0,-0.5) -- (0.0,0.05);
      \draw[boson] (0,-0.5) -- (-0.5,-1.5);
      \draw[boson] (0.2,-1.0) -- (0.5,-1.5);
      \draw[boson] (0.2,-1.0) -- (0.0,-1.5);
      \draw[scalar] (0.2,-1.0) -- (0,-0.5);
      \draw [fill] (-.5,-1.5) circle [radius=2pt];
       \draw [fill] (0.0,-1.5) circle [radius=2pt];
         \draw [fill] (0.5,-1.5) circle [radius=2pt];
      \node[label=below:\small{$J^{(1)}$}] at (-0.5,-1.5) {};
      \node[label=below:\small{$J^{(1)}$}] at (0.5,-1.5) {};
       \node[label=below:\small{$J^{(1)}$}] at (0.0,-1.5) {};
    \end{tikzpicture}\right)
\end{equation}
which can be expressed using the lowest-order result as
\begin{equation}
  \label{e:Y3result}
    Y^{a\ (3)}_{bc}=\begin{tikzpicture}[baseline={([yshift=-.5ex]current bounding box.center)},scale=1]
      \draw[scalar] (0,-0.5) -- (0.0,0.05);
      \draw[boson] (0,-0.5) -- (-0.5,-1.5);
      \draw[boson] (0,-0.5) -- (0.5,-1.5);
      \draw [fill] (-.5,-1.5) circle [radius=2pt];
         \draw [fill] (0.5,-1.5) circle [radius=2pt];
      \node[label=below:\small{$J^{(1)}$}] at (-0.5,-1.5) {};
      \node[label=below:\small{$J^{(2)}$}] at (0.5,-1.5) {};
    \end{tikzpicture}+\begin{tikzpicture}[baseline={([yshift=-.5ex]current bounding box.center)},scale=1]
      \draw[scalar] (0,-0.5) -- (0.0,0.05);
      \draw[boson] (0,-0.5) -- (-0.5,-1.5);
      \draw[scalar] (0,-0.5) -- (0.5,-1.5);
      \draw [fill] (-.5,-1.5) circle [radius=2pt];
         \draw [fill] (0.5,-1.5) circle [radius=2pt];
      \node[label=below:\small{$J^{(1)}$}] at (-0.5,-1.5) {};
      \node[label=below:\small{$Y^{(2)}$}] at (0.5,-1.5) {};
    \end{tikzpicture}.
\end{equation}
  
\subsection{All-loop order}
\label{sec:all-loop}

The previous iterative structure carries over to all loop  orders. The
fact that the auxiliary field does not couple directly to the massive
line and that there are only cubic interactions leads to 
the expression of the  $(n-1)$-loops graviton emission as a sum of
lowest-loop contributions
\begin{equation}\label{e:Jnrecursion}
J^{(n)}=
    \sum_{m=1}^{n-1}\left(
    \begin{tikzpicture}[baseline={([yshift=-.5ex]current bounding box.center)},scale=1]
      \draw[boson] (0,-0.5) -- (0.0,0.05);
      \draw[boson] (0,-0.5) -- (-0.5,-1.5);
      \draw[boson] (0,-0.5) -- (0.5,-1.5);
      \draw [fill] (-.5,-1.5) circle [radius=2pt];
         \draw [fill] (0.5,-1.5) circle [radius=2pt];
      \node[label=below:\small{$J^{(m)}$}] at (-0.5,-1.5) {};
      \node[label=below:\small{$J^{(n-m)}$}] at (0.5,-1.5) {};
    \end{tikzpicture}-
     \begin{tikzpicture}[baseline={([yshift=-.5ex]current bounding box.center)},scale=1]
      \draw[boson] (0,-0.5) -- (0.0,0.05);
      \draw[boson] (0,-0.5) -- (-0.5,-1.5);
      \draw[scalar] (0,-0.5) -- (0.5,-1.5);
      \draw [fill] (-.5,-1.5) circle [radius=2pt];
         \draw [fill] (0.5,-1.5) circle [radius=2pt];
      \node[label=below:\small{$J^{(m)}$}] at (-0.5,-1.5) {};
      \node[label=below:\small{$Y^{(n-m)}$}] at (0.5,-1.5) {};
    \end{tikzpicture}-
    \begin{tikzpicture}[baseline={([yshift=-.5ex]current bounding box.center)},scale=1]
      \draw[boson] (0,-0.5) -- (0.0,0.05);
      \draw[scalar] (0,-0.5) -- (-0.5,-1.5);
      \draw[scalar] (0,-0.5) -- (0.5,-1.5);
      \draw [fill] (-.5,-1.5) circle [radius=2pt];
         \draw [fill] (0.5,-1.5) circle [radius=2pt];
      \node[label=below:\small{$Y^{(m)}$}] at (-0.5,-1.5) {};
      \node[label=below:\small{$Y^{(n-m)}$}] at (0.5,-1.5) {};
    \end{tikzpicture}\right)
\end{equation}
and, similarly for the emission of the auxiliary field
\begin{equation}\label{e:Ynrecursion}
    Y^{(n)}=\sum_{m=1}^{n-1}\left(
    \begin{tikzpicture}[baseline={([yshift=-.5ex]current bounding box.center)},scale=1]
      \draw[scalar] (0,-0.5) -- (0.0,0.05);
      \draw[boson] (0,-0.5) -- (-0.5,-1.5);
      \draw[boson] (0,-0.5) -- (0.5,-1.5);
      \draw [fill] (-.5,-1.5) circle [radius=2pt];
         \draw [fill] (0.5,-1.5) circle [radius=2pt];
      \node[label=below:\small{$J^{(m)}$}] at (-0.5,-1.5) {};
      \node[label=below:\small{$J^{(n-m)}$}] at (0.5,-1.5) {};
    \end{tikzpicture}+
     \begin{tikzpicture}[baseline={([yshift=-.5ex]current bounding box.center)},scale=1]
      \draw[scalar] (0,-0.5) -- (0.0,0.05);
      \draw[boson] (0,-0.5) -- (-0.5,-1.5);
      \draw[scalar] (0,-0.5) -- (0.5,-1.5);
      \draw [fill] (-.5,-1.5) circle [radius=2pt];
         \draw [fill] (0.5,-1.5) circle [radius=2pt];
      \node[label=below:\small{$J^{(m)}$}] at (-0.5,-1.5) {};
      \node[label=below:\small{$Y^{(n-m)}$}] at (0.5,-1.5) {};
    \end{tikzpicture}\right).
\end{equation}

The $n-1$-loop contribution is therefore given by a finite sum of
generalised one-loop diagrams with effective vertices given by
the lowest-loop contributions. These iterated diagrams evaluate directly into the multiloop massless bubble master integrals of Sec.~2.2 of~\cite{Mougiakakos:2020laz}. The fact that the classical result \emph{is}
only built from binary tree graphs is an important simplification over previous amplitude approaches~\cite{Boulware:1968zz,Duff:1973zz,Bjerrum-Bohr:2002fji,Neill:2013wsa,Jakobsen:2020ksu,Mougiakakos:2020laz,DOnofrio:2022cvn} that involved either higher graviton vertices or non-minimal couplings. 

In the perturbiner method~\cite{Damgaard:2024fqj}, a similar recursion relation is derived for the off-shell currents of the \emph{inverse gothic}. It is interesting to note that both our amplitude based and the perturbiner method approach utilize the harmonic gauge and introduce an auxiliary DOF~\eqref{e:Adef}
in our case and  $\tilde{\mathfrak{g}}_{ab}=\eta_{ab}+\tilde{h}_{ab}$ in~\cite{Damgaard:2024fqj}. Furthermore, the specific choice of the auxiliary DOF is crucial for the derivation of the recursion relations in each approach. It is apparent that the choice of suitable gauge and DOF is formalism dependent and there is an optimal choice corresponding to each formalism. This observation is further supported by the results of~\cite{Mougiakakos:2020laz}, where it was not possible to derive the metric to all orders in perturbation theory working with the metric perturbation $g_{\mu\nu}=\eta_{\mu\nu}+h_{\mu\nu}$ as a DOF only.

\section{Form factors recursion relations}
From the iterative structure of the diagrammatic expansion we can
derive recursion relations for the form factors, in general-$D$ given
by the bilinear recursion relation
for $1\leq k\leq 8$
\begin{equation}\label{value}
\chi_k^{(n)}(D)=\sum_{i,j=1}^8\sum_{m=1}^{n-1}\chi_i^{(m)}(D)\chi_j^{(n-m)}(D)M_k^{ij}(D)\,
,
\end{equation}
together with the initial conditions provided by the tree-level result
in~\eqref{e:val1}.
We have explicitly verified to any order and any dimension-$D$ that
$\chi_2^{(n)}+\chi_3^{(n)}=0$, as required from the conservation of
the effective stress-tensor/off-shell current
$k^{\mu}h^{(n)}_{\mu\nu}(\textbf{x})=k^{\mu}J^{(n)}_{\mu\nu}(\textbf{k})=0$. The matrix coefficients
$M^k_{ij}(D)$ for $1\leq k\leq 8$ are provided  in the Appendix and on the repository~\cite{githubmetric}.

After Fourier transforming to position space we can write the classical solution with respect to the \textit{form factors} as, using  $\rho(r,D)$  introduced in~\eqref{e:rhodef},
\begin{multline}\label{e:hresult}
h^{(n)}_{\mu\nu}(\textbf{x})=\rho(r,D)^n\Big[\left(\chi_1^{(n)}-\chi_2^{(n)}\right)\delta_{\mu}^0\delta_{\nu}^0\cr+\chi_2^{(n)}\frac{1-(n-1)(D-3)}{2-(n-1)(D-3)}\,\delta_{ij}\cr+\chi_2^{(n)}\frac{n(D-3)}{2-(n-1)(D-3)}\,
n_\mu n_\nu\Big]
\end{multline}
and for the auxiliary field
\begin{multline}\label{e:Aresult}
A^{a\ (n)}_{bc}(\textbf{x})=\rho(r,D)^n\frac{n(D-3)}{r}\Bigg[\left(\chi_8^{(n)}-\chi_7^{(n)}\right)\delta^a_0n_{(b}\delta_{c)}^0\cr+\left(\chi_4^{(n)}-\chi_5^{(n)}\right)n^a\delta_b^0\delta_c^0\cr+\left(\chi_5^{(n)}-\frac{\chi_6^{(n)}}{(n-1)(D-3)-2}\right)n^a\delta_{bj}\delta_{ck}
                      \cr      +\chi_6^{(n)}\frac{n(D-3)+2}{(n-1)(D-3)-2}\,1n^an_bn_c
                            \cr+\left(\chi_8
                              ^{(n)}-\frac{\chi_6^{(n)}}{(n-1)(D-3)-2}\right)\delta^{ai}\delta_{j(b}n_{c)}\Bigg]
\end{multline}

\section{Schwarzschild black-hole in $D=4$}

In $D=4$ the solution to the recursion relations of the form factors
in~\eqref{e:val2}
\begin{equation}
    \boldsymbol{\mathbf{\chi}}^{(2)}(4)=\left({15\over2},{1\over2},-{1\over2},6,2,0,-4,0\right)
\end{equation}
and~\eqref{value} for $n\ge 3$ give
\begin{multline}\label{e:valn}
  \boldsymbol{\chi}^{(n)}(4)=\Big(8, 0, 0,
  4+n(-1)^n+\frac{1+3(-1)^n}{2n(n+2)}, \cr 
  2+\frac{1+3(-1)^n}{2n(n+2)}, \frac{1+3(-1)^n}{2n(n+2)}(n-3), \cr \frac{1}{n}-4-\frac{1+3(-1)^n}{2n(n+2)}(n+1), \frac{1+3(-1)^n}{2n(n+2)} \Big) 
\end{multline}
which plugged in~\eqref{e:hresult},~\eqref{e:Aresult} with $\rho=\rho(r,4)=G_Nm/r$ and $n_i=x_i/r$ reproduce exactly the classical solutions in~\eqref{e:hresult.class},~\eqref{e:Aresult.class}.

The present derivation does not need introducing non-minimal couplings~\cite{Goldberger:2004jt} for removing the
ultraviolet divergences that occurred in the previous perturbative
computations~\cite{Mougiakakos:2020laz,DOnofrio:2022cvn}. 
The momentum space  expressions~\eqref{e:Jansatz}
and~\eqref{e:Yansatz} are divergent for $n=\frac{2m+(D-1)}{D-3},\ n,m \in \mathbb{N}$, since the
$\rho(|k|,D,n)$ develops a pole, which in $D=4$ corresponds to two-loop
divergence of~\cite{Mougiakakos:2020laz,DOnofrio:2022cvn}. But this momentum space
divergence is cancelled after Fourier transforming to direct space, thanks to the parametrization of the off-shell currents,
and the final expressions~\eqref{e:hresult} and~\eqref{e:Aresult} are
free of divergences  because 
the form factors $\boldsymbol{\chi}^{(n)}(D)$ are finite for $D=4$.

\section{Discussion}
We have succesfully derived the exact classical black-hole solutions~\eqref{e:hresult.class} and~\eqref{e:Aresult.class}
generated by a massive source to all orders in $G_N$ in four dimensions.  The crucial ingredient of
our approach is the cubic formulation~\cite{Cheung:2017kzx} of general
relativity combined with the powerful worldline EFT formalism.  Specifically, we were able to organize the
computation of the \textit{form factors} from the effective off-shell
currents for the graviton and auxiliary field emission to any order in
$G_N$, into recursion relations relating the various order in
perturbation. These recursion relations originate from the simple
diagrammatic expansion given by binary trees in~\eqref{e:Jnrecursion} and~\eqref{e:Ynrecursion}. 

The computation presented in this paper shows that with a suitable choice of gauge and DOF, amplitude based classical GR computations can be simplified immensely. We can find a closed form for the \emph{form factors} in $D=4$ because within our setup all the residual freedom of harmonic coordinates  within the harmonic gauge  is fixed~\cite{Fromholz:2013hka}. Most importantly,  the  all-order derivation is possible because no  non-minimal worldline couplings nor quartic or higher order self-interaction vertices are needed, rendering the computation maximally minimal from an amplitude point of view. The form factors $\boldsymbol{\chi}^{(n)}(D)$ have been determined in general dimensions, but in order to establish a connection with the Schwarzschild-Tangherlini metric, it is necessary to match the gauge used in the amplitude computation with the gauge used in the GR side.  A mapping can be determined perturbatively in powers of $\rho(r,D)$, but it is difficult to find an all order expression.

The knowledge of the exact metric generated by a massive source is a first step toward  an amplitude based approach~\cite{Kosmopoulos:2023bwc,Cheung:2023lnj,Adamo:2023cfp,Driesse:2024xad} to the self-force calculation~\cite{Barack:2018yvs}. 
We expect that the simplicity of the cubic interactions, presented in this work, will be instrumental to setup a complete self-force calculation from an amplitude calculation.
 Furthermore, the formalism presented here could be extended to include spin contributions, since the simplicity of the computation resulting from the restriction to cubic interactions would not be spoiled in this case as well.

 \section{Acknowledgements}
We would like to thank Poul Damgaard, Leonardo de la Cruz and Donal O'Connell for useful comments and discussions.  The work of P.V. has received funding from the ANR grant ``SMAGP'' ANR-20-CE40-0026-01. S.M. acknowledges financial support by ANR PRoGRAM project, grant ANR-21-CE31-0003-001.

\appendix
\section{Feynman rules}

 Restoring the normalization factors give an extra $1/2$ in the
 propagators and $2\sqrt{32\pi G_N}$ to the three-point vertices compared to~\cite{Cheung:2017kzx}. 

\noindent {\bf The graviton propagator}
\begin{equation}\label{e:PropG}
\mathcal{P}^h_{\alpha\beta\mu\nu}=-\frac{i}{2}\,\frac{\eta_{\mu\alpha}\eta_{\nu\beta}+\eta_{\nu\alpha}\eta_{\mu\beta}-\eta_{\mu\nu}\eta_{\alpha\beta}}{k^2}.
\end{equation}
Notice that the numerator does not depend on the space-time dimension
and does not have the usual $2/(D-2)$ factor multiplying the
$\eta_{\mu\nu}\eta_{\alpha\beta}$ term~\cite{Veltman:1975vx}, needed
for propagating only the transverse degree-of-freedom of the
graviton. This is compensated by the $D$ dependence in the vertices
derived from the world-line action.

\noindent{\bf
The auxiliary field propagator}
\begin{equation}
\mathcal{P}^{A\ a \ d}_{\ \ \  \ bc\ ef}=-\frac{i}{2}\,\frac{1}{k^2}\left(\delta^{a}_{f}\delta^{d}_{c}\eta_{be}+\frac{\eta^{ad}}{2}\left(\frac{\eta_{bc}\eta_{ef}}{D-2}-\eta_{be}\eta_{cf}\right)\right).
\end{equation}

\noindent{\bf
The three-graviton vertex}
\begin{multline}
V_{h^3}^{abcdef}(p_1,p_2,p_3)=2i\sqrt{32\pi G_N}\,\cr
\times\Bigg{\{}\Bigg[
\frac{1}{2}\left(\frac{\eta^{ab}\eta^{cd}}{D-2}-\eta^{ac}\eta^{bd}\right)p_{1}^ep_{2}^f-p_1^cp_2^a\eta^{be}\eta^{fd}\cr
+
\left(\eta^{ac}\eta^{de}\eta^{fb}
                                                                    -\frac{1}{2(D-2)}\left(\eta^{ab}\eta^{ce}\eta^{fd}+\eta^{cd}\eta^{ae}\eta^{fb}\right)\right)(p_1\cdot p_2)
\Bigg]\cr
+\Bigg[\begin{matrix}
    p_2\leftrightarrow p_3 \\
    cd\leftrightarrow ef 
\end{matrix}\Bigg]+\Bigg[\begin{matrix}
    p_1\leftrightarrow p_3 \\
    ab\leftrightarrow ef 
\end{matrix}\Bigg]\Bigg{\}}.
\end{multline}

\noindent{\bf
The two-graviton and one auxiliary field vertex}
\begin{multline}
V^{abcd\ fg}_{h^2A\ \ \ e}(p_1,p_2,p_3)=2\sqrt{32\pi G_N}\,\cr
\times\Bigg{\{}\Bigg[p_1^d\eta^{cf}\left(\delta^a_e\eta^{bg}-\frac{\delta^g_e\eta^{ab}}{D-2}\right)-p_1^g\eta^{cf}\left(\delta^a_e\eta^{bd}-\frac{\delta^d_e\eta^{ab}}{D-2}\right) \Bigg] \cr
+\Bigg[\begin{matrix}
    p_1\leftrightarrow p_2 \\
    ab\leftrightarrow cd 
\end{matrix}\Bigg]-\eta^{fa}\eta^{bc}\eta^{dg}p_{3e}\Bigg{\}}.
\end{multline}

\noindent{\bf
The one-graviton two auxiliary fields vertex}
\begin{multline}
V^{ab\ \ de\ \ gh}_{hA^2\ \ \ c\ \ \ f}(p_1,p_2,p_3)=2i\sqrt{32\pi G_N}\cr
\times\Bigg{\{}2\eta^{ah}\eta^{bd}\left(\delta^{g}_{c}\delta^{e}_{f}-\frac{\delta^{e}_{c}\delta^{g}_{f}}{D-1}\right)\Bigg{\}}.
\end{multline}

In the main text, we also use the normalized Feynman rules denoted with a bar as
\begin{align}
&\mathcal{P}^h_{\alpha\beta\mu\nu}=-\frac{i}{2}\,\bar{\mathcal{P}}^h_{\alpha\beta\mu\nu},\cr
&\mathcal{P}^{A\ a \ d}_{\ \ \  \ bc\ ef}=-\frac{i}{2}\,\bar{\mathcal{P}}^{A\ a \ d}_{\ \ \  \ bc\ ef},\cr
&V_{h^3}^{abcdef}=2i\sqrt{32\pi G_N}\,\bar{V}_{h^3}^{abcdef},\cr
&V^{abcd\ fg}_{h^2A\ \ \ e}=2\sqrt{32\pi G_N}\,\bar{V}^{abcd\ fg}_{h^2A\ \ \ e},\cr
&V^{ab\ \ de\ \ gh}_{hA^2\ \ \ c\ \ \ f}=2i\sqrt{32\pi G_N}\,\bar{V}^{ab\ \ de\ \ gh}_{hA^2\ \ \ c\ \ \ f}.\label{e:BarPV}
\end{align}

\section{The recursion relation}\label{sec:recursion}
We explicitly give the matrices $M^k(D)$ of the  recursion relations in~\eqref{value}. 
The matrices take the block diagonal form
\begin{equation}
  M_k(D)=
  \begin{pmatrix}
    m_k^1&0&0\cr
    0 &0& m_k^2
  \end{pmatrix}
\end{equation}
where $m^1_k$ are $3\times 3$ upper-triangle matrices and $m_2^k$ are
upper-triangular $5\times 4$ matrices with $m^2_r=0$ for $4\leq r\leq
8$.
The recursion relation takes the form for $1\leq i\leq 3$
\begin{multline}
  \chi^{(n)}_l=\sum_{m=1}^{n-1}\sum_{r,s=1}^3 \chi^{(m)}_r(D)
                (m^1_l)^{rs}(D) \chi_s^{(n-m)}(D)\cr +\sum_{m=1}^{n-1}\sum_{r,s=4}^8 \chi^{(m)}_r(D)
                (m^2_l)^{rs}(D) \chi_s^{(n-m)}(D)
              \end{multline}
              and for $4\leq i\leq 8$
\begin{equation}
  \chi^{(n)}_l=\sum_{m=1}^{n-1}\sum_{r,s=1}^3 \chi^{(m)}_r(D)
                (m^1_l)^{rs}(D) \chi_s^{(n-m)}(D).
              \end{equation}
In $D=4$ dimensions the matrices are given by [general dimension
expressions are in the file \texttt{Matrices-recursion.txt} on the repository~\cite{githubmetric}]
\begin{widetext}
\begin{equation}
  m_1^1=
  \begin{pmatrix}
   \frac{m^2 (4 n+5)-m n (4 n+5)+5 n \left(n^2-1\right)}{6 n \left(n^2-1\right)} & -\frac{m^2 n+m \left(n^2-2\right)+n^3-n}{6 n \left(n^2-1\right)} & -\frac{-\left(m^3
   (n+1)\right)+m^2 (n-1)+m (n (n+2)-1)+(n-2) (n-1) n (n+1)}{6 n \left(n^2-1\right) (m-n+2)} \\
 0 & \frac{m (m-n)}{3 n-3 n^3} & \frac{m \left(3 m^2+m (n-1)-5 n^2+n+1\right)}{12 n \left(n^2-1\right) (m-n+2)} \\
 0 & 0 & \frac{8 m^4+m^2-2 (m+2) n^3+(m+2) (10 m-1) n^2-(m (16
   m+19)+1) m n+2 n^4+4 n}{24 (m-2) n \left(n^2-1\right) (m-n+2)}
  \end{pmatrix}
\end{equation}
\begin{equation}
{\tiny m_2^1=  \begin{pmatrix}
     \frac{m^2 (n+2)-m (n+2) n+n^3-n}{6 n \left(n^2-1\right)} & \frac{(m-n) \left(m-n^2+1\right)}{6 n \left(n^2-1\right)} & \frac{m^3 (n+1)-2 m^2 \left(n^2+n-2\right)+m
   \left(2 n^3-5 n+1\right)-(n-2) (n-1) n (n+1)}{12 n \left(n^2-1\right) (m-n+2)} \\
 0 & \frac{-m^2 (2 n+3)+m (2 n+3) n-3 n^3+3 n}{3 n \left(n^2-1\right)} & \frac{-3 m^3 (2 n+1)+m^2 \left(16 n^2+n-17\right)+m (n ((9-16 n) n+23)-7)+6 (n-2) (n-1) n
   (n+1)}{12 n \left(n^2-1\right) (m-n+2)} \\
 0 & 0 & \frac{-8 m^4 n+16 m^3 n^2+m^2 (n (5-n (16 n+5))+24)+m n (n (n (8 n+5)-5)-24)-5 (n-2) (n-1) n (n+1)}{24 (m-2) n \left(n^2-1\right) (m-n+2)}
  \end{pmatrix}}
\end{equation}
\begin{equation}
  m_3^1=
  \begin{pmatrix}
     \frac{m (m-n)}{3 n (n+1)} & \frac{m (m-n)}{3 n (n+1)} & \frac{m (n-3) (n-m)}{6 n (n+1) (m-n+2)} \\
 0 & \frac{2 m (n-m)}{3 n (n+1)}+1 & \frac{(4 m+3) n^2+2 ((3-2 m) m+3) n+3 m ((m-2) m-1)-3 n^3}{6 n (n+1) (m-n+2)} \\
 0 & 0 & \frac{8 m^4-16 m^3 n+m^2 (n (16 n-7)+7)+m n ((7-8 n) n-7)+3 (n-2) n (n+1)}{12 (m-2) n (n+1) (m-n+2)} 
  \end{pmatrix}
\end{equation}
\begin{equation}
  m_4^1=
  \begin{pmatrix}
    \frac{5}{12} & -\frac{2 m+n}{12 n} & \frac{m^2 (4 n+3)-m (n (2 n+3)+2)+2 n \left(-n^2+n+2\right)}{24 n (n+1) (m-n+2)} \\
 0 & 0 & \frac{m (2 m-3 n-1)}{24 n (n+1) (m-n+2)} \\
 0 & 0 & \frac{-5 m^2+5 m n+(n-2) n (n+1)}{24 (m-2) n (n+1) (m-n+2)}
  \end{pmatrix};\quad
 m_5^1=
 \begin{pmatrix}
    \frac{1}{12} & -\frac{1}{12} & \frac{m (2 n+1) (m-n)}{24 n (n+1) (m-n+2)} \\
 0 & \frac{1}{6} & \frac{-2 m^2 (2 n+1)+m \left(6 n^2+n-3\right)+2 n \left(-n^2+n+2\right)}{24 n (n+1) (m-n+2)} \\
 0 & 0 & \frac{m^2 (6 n+1)-m n (6 n+1)+n \left(n^2-n-2\right)}{24 (m-2) n (n+1) (m-n+2)} 
 \end{pmatrix}
\end{equation}
\begin{equation}
m_6^1=
\begin{pmatrix}
   -\frac{1}{4} & \frac{1}{4} & \frac{m (n-m)}{24 n (n+1) (m-n+2)} \\
 0 & 0 & \frac{m (2 m+n+3)}{24 n (n+1) (m-n+2)} \\
 0 & 0 & \frac{m (4 n-1) (m-n)}{24 (m-2) n (n+1) (m-n+2)}
\end{pmatrix};
m_7^1=
\begin{pmatrix}
   0 & 0 & \frac{m (n-m)}{24 n (n+1) (m-n+2)} \\
 0 & \frac{1}{4} & \frac{m (2 m+n+3)}{24 n (n+1) (m-n+2)} \\
 0 & 0 & \frac{m (4 n-1) (m-n)}{24 (m-2) n (n+1) (m-n+2)}
\end{pmatrix};
m_8^1=
\begin{pmatrix}
   0 & 0 & -\frac{m (n-2) (m-n)}{12 n (n+1) (m-n+2)} \\
 0 & 0 & \frac{(n-2) (m-n) (2 m+3 n+3)}{12 n (n+1) (m-n+2)} \\
 0 & 0 & \frac{(n-2) \left(m^2 (8 n-2)+2 m (1-4 n) n+3 n (n+1)\right)}{24 (m-2) n (n+1) (m-n+2)} 
\end{pmatrix}
\end{equation}

\begin{equation}
  m_1^2=
  \begin{pmatrix}
   0 & \frac{2 m (n-m)}{(n-1) n} & \frac{2 m (m-n)}{(n-1) n} & 0 \\
 \frac{3 m (m-n)}{4 n \left(n^2-1\right)} & \frac{m (8 n+9) (m-n)}{4 n \left(n^2-1\right)} & \frac{m (m-n)}{2 n-2 n^3} & \frac{m (m-n) (3 m+n+2)}{4 n \left(n^2-1\right)
   (m-n+2)} \\
 0 & \frac{3 m (m-n)}{4 n \left(n^2-1\right)} & \frac{m (m-n)}{2 n \left(n^2-1\right)} & \frac{m (m-n)}{4 n \left(n^2-1\right)} \\
 0 & 0 & \frac{m (m-n)}{n-n^3} & \frac{m (m-n) (m+n)}{2 n \left(n^2-1\right) (m-n+2)} \\
 0 & 0 & 0 & \frac{m (3 m (m-n)-4 n-2) (m-n)}{4 (m-2) n \left(n^2-1\right) (m-n+2)}
  \end{pmatrix}
\end{equation}
\begin{equation}
m_2^2=  \begin{pmatrix}
   0 & \frac{m (n-m)}{(n-1) n} & \frac{m (m-n)}{(n-1) n} & 0 \\
 \frac{3 m (n-m)}{4 \left(n^2-1\right)} & \frac{m (3 n+4) (m-n)}{4 n \left(n^2-1\right)} & \frac{m (5 n+6) (n-m)}{2 n \left(n^2-1\right)} & -\frac{m (m-n) (n (3 m-3
   n+2)+4)}{4 n \left(n^2-1\right) (m-n+2)} \\
 0 & \frac{3 m (n-m)}{4 \left(n^2-1\right)} & \frac{m (n-m)}{2 \left(n^2-1\right)} & \frac{m (n-m)}{4 \left(n^2-1\right)} \\
 0 & 0 & \frac{m (m-n)}{n^2-1} & -\frac{m (m-n) (n (m-n)+2)}{2 n \left(n^2-1\right) (m-n+2)} \\
 0 & 0 & 0 & -\frac{m (m-n) (n (3 m (m-n)-2)-4)}{4 (m-2) n \left(n^2-1\right) (m-n+2)}
  \end{pmatrix}
\end{equation}
\begin{equation}
  m_3^2=
  \begin{pmatrix}
  0 & 0 & 0 & 0 \\
 \frac{3 m (m-n)}{2 n (n+1)} & \frac{m (m-n)}{2 n (n+1)} & \frac{m (n-m)}{n (n+1)} & \frac{m (3 m-3 n-2) (m-n)}{2 n (n+1) (m-n+2)} \\
 0 & \frac{3 m (m-n)}{2 n (n+1)} & \frac{m^2-m n}{n^2+n} & \frac{m (m-n)}{2 n (n+1)} \\
 0 & 0 & \frac{2 m (n-m)}{n (n+1)} & \frac{m (m-n-2) (m-n)}{n (n+1) (m-n+2)} \\
 0 & 0 & 0 & \frac{m (3 m (m-n)+2) (m-n)}{2 (m-2) n (n+1) (m-n+2)}
  \end{pmatrix}
\end{equation}
\end{widetext}

\end{document}